Structural Heterogeneity, Ductility, and Glass Forming Ability of Zr-Based Metallic Glasses

Soohyun Im[1], Pengyang Zhao[1,2], Geun Hee Yoo[3], Zhen Chen[4], Gabriel Calderon[1], Mehrdad Abbasi Gharacheh[1], Olivia Licata[5], Baishakhi Mazumder[5], David A. Muller[4,6], Eun Soo Park[3], Yunzhi Wang[1], and Jinwoo Hwang[1, *]

1. Department of Materials Science and Engineering, The Ohio State University, Columbus, OH 43212, USA
2. Department of Engineering Mechanics, Shanghai Jiao Tong University, 800 Dongchuan Road, Shanghai, 200240, China
3. Research Institute of Advanced Materials, Department of Materials Science and Engineering, Seoul National University, Seoul 08826, South Korea
4. School of Applied and Engineering Physics, Cornell University, Ithaca, NY 14853, USA
5. Department of Materials Design and Innovation, University at Buffalo-SUNY, Buffalo, NY, USA
6. Kavli Institute at Cornell for Nanoscale Science, Cornell University, Ithaca, NY 14853, USA

* Corresponding author

Abstract

We show the correlation between nanoscale structural heterogeneity and mechanical property and glass forming ability of Zr-based metallic glasses (MGs). Detailed parameters of medium range ordering (MRO) that constitutes the structural heterogeneity, including the type, size, and volume fraction of MRO domains determined using 4-dimensional scanning transmission electron microscopy, directly correlate with the ductility and glass forming ability of Zr-Cu-Co-Al MGs. Mesoscale deformation simulation incorporating the experimentally determined MRO confirms that the diverse types and sizes of MRO can significantly influence the MGs' mechanical behavior.

Different from crystalline solids where extended defects such as dislocations and grain boundaries are well defined and characterized, identifying, charactering and describing extended defects in amorphous solids where long range atomic orders are absent are still semantically complex and experimentally and computationally challenging. The same is true for our adventure in understanding and establishing the microstructure-property relationships in these two distinctively different classes of solids. For example, short-range ordering (SRO) within a few nearest neighbor shells (typically less than ~ 1 nm in length) in metallic glasses (MGs) has been extensively investigated using, inter alia, large area diffraction and pair distribution functions (PDFs), in attempts to uncover connection to many important properties [1–3]. PDFs, however, rely on an inherent averaging of structures over the probed area, making it challenging to study local heterogeneity of ordering at the nanometer scale. The nanoscale heterogeneity, also commonly referred as medium-range ordering (MRO), has been under extensive scrutiny recently. Atomistic models have provided valuable insights on the possible structures of MRO domains [4–6]. Due to the time scale limits of these models, for example, the extremely fast quenching rates in molecular dynamics (MD) simulations, however, it is currently unclear whether the simulations have generated realistic MRO structures comparable to those found in real MGs. Meanwhile, the studies involving electron diffraction and fluctuation microscopy [7–15] have provided new ways to characterize MRO and local heterogeneity. The heterogeneity in properties (*e.g.* elastic modulus) has also been measured using dynamic force microcopy [16], although the property may be limited to the surface and its relationship to structural heterogeneity (*i.e.* MRO) remains unclear at this point.

MGs can display a wide range of ductilities: while most MGs are brittle, some MGs have shown dramatic increases in ductility (matching that of ductile crystalline materials) with only a small change in their compositions [17–19]. However, the exact mechanism of this remarkable effect has yet to be identified. Free volume has been frequently used to explain the ductility of MGs [20], but in many cases, they do not appear to have any direct connection to the ductility or other properties [21,22], making it difficult to use the free volume argument alone in general. Recent theories have suggested more advanced concepts, such as the flexibility volume that may be associated with geometrically unfavored motifs (GUMs) that are deviated from the stable icosahedral SRO [23,24]. This new insight is valuable as they have evolved from the free volume theory and endorses the important viewpoint that some local structural heterogeneity (which exists in as-quenched MGs [9]) may be closely related to their deformation behavior. However, since the simulations impose much higher quenching rates than those in real experiments, it is still unclear whether the same population of icosahedral or GUM clusters are present in real MGs.

The important question is then whether the MRO domain structures of real MGs vary with composition and processing history and, if so, how its impact is on the ductility and overall deformation behavior of MGs. Small changes in composition typically do not lead to significant changes in PDF, suggesting that the change in structure by composition, if any, may be localized at the nanometer scale. If so, such nanoscale details could be lost in PDF due to the inherent averaging process. The potential involvement of nanoscale heterogeneity, or MRO, in plastic deformation of MGs seems to coincide with the shear transformation zone (STZ) proposed in theory of MG deformation [20], where the STZs have their length scale similar to that of MRO [20,25,26]. If the correlation between MRO and plastic deformation in MGs could be established, it will lead to significant advances in MG design based on controlling MRO populations and activities through alloying and processing. Some MD simulations have suggested that STZ activities may be correlated (at least statistically) with local atomic ordering [23]. In addition, the new information on MRO will be crucial to connect local atomic structure to glass forming ability of MGs that has been under extensive investigation (*e.g.* [2,5]) and to the current understanding of structural evolution in MGs (*e.g.* relaxation or rejuvenation) (*e.g.* [27]).

In this paper, we experimentally determine the MRO structure in detail and correlate it to the ductility and glass forming ability of Zr-based MGs. Our MRO characterization is based on a 4-dimensional scanning transmission electron microscopy (4D-STEM) technique, utilizing the quantitative analysis of data acquired using the electron microscopy pixel array detector (EMPAD) with high dynamic range [28,29]. 4D-STEM reveals a comprehensive map of local structural heterogeneity that provides direct information about MRO, including the type (symmetry and composition), size distribution, spatial distribution, and volume (or area) fraction of MRO domains with high statistical precision. We show that these detailed MRO parameters directly correlate with the changes in ductility and glass forming ability of Zr-Cu-Co-Al MGs. We then use mesoscale deformation simulation that directly incorporates the experimentally determined MRO information to show the correlation between the deformation mechanism and MRO structure [30–32]. Our simulation can flexibly assume and integrate the important characteristics of MRO domains to study their impact on the overall deformation behavior beyond the spatial and temporal limits of atomistic simulations. The simulation reveals that the degree of heterogeneity in MRO domain structures, both in terms of type and size, directly correlates to the ductility of the MGs.

Our 4D-STEM analysis is based on the fast acquisition of electron nanodiffraction patterns in 2D reciprocal space ($k_x$, $k_y$) throughout many MG sample areas in 2D real space ($x$, y) with spatial oversampling of electron probes with a diameter of 1.0 nm [28] (Fig. 1a). The signal was recorded using an EMPAD with high dynamic range (32 bit) that is essential for the quantitative analysis of the signal [29]. About 250,000 nanodiffraction patterns were acquired per sample. The acquired nanodiffraction patterns were reconstructed in the real space using each ($k_x$, $k_y$) position to reveal the detailed maps of MRO, with each map size of 40 × 40 nm$^2$ for the entire $\varphi$ range (Fig. 1a) and the $k$ range up to ~ 6 nm$^{-1}$. The example maps for $k$ = 4 nm$^{-1}$ with different $\varphi$ are shown in Fig. 1b and 1c. These maps show the nanoscale speckles with high intensity, which are the electron intensities directly scattered by the local MRO regions toward that particular $k$ and $\varphi$. We then determine the size and area fraction of MRO speckles within those maps for the entire $k$ range. $k$ is inverse of the 'interplanar spacing' within the MRO and therefore related to the type of MRO. Only the areas with sample thickness of ~ 25 - 35 nm was used for the analysis to prevent any complication, such as the effect of plural scattering [13,28,33]. The details of the quantitative determination of the MRO size as well as thickness filtering are described in Supplemental Materials. To achieve reliable statistics, the information was averaged over 126 maps acquired using different $\varphi$, per $k$ and per sample area. Same process was repeated over 4 areas per sample, and then results were averaged over those areas.

Other experimental details, including MG sample preparation, TEM sample preparation, and atom probe tomography, are described in Supplemental Materials.

The Zr-Cu-Co-Al MG system (Fig. 2a) that we investigated is essentially a mixture of two glass forming systems [19], i.e., $(Zr_{45}Cu_{50}Al_5)_{1-x}$ and $(Zr_{55}Co_{25}Al_{20})_x$. When $x = 0.5$, which is $Zr_{50}Cu_{25}Co_{12.5}Al_{12.5}$, the glass shows substantial increase in ductility as compared to $Zr_{45}Cu_{50}Al_5$ ($x = 0$) and $Zr_{55}Co_{25}Al_{20}$ ($x = 1$) (Fig. 2b) (also see [19]). Differential scanning calorimetry (DSC) shows multiple crystallization peaks (Fig. 2c), suggesting that there may be phase separation occurring at $x = 0.5$ (also see [19]). However, frequency distribution and nearest neighbor distribution analyses of the atom probe tomography data (Fig. 2d) did not detect any segregation or clustering at $x = 0.5$. This suggests that if any structural change or chemical segregation occurs in association with the composition, it must occur at a small scale, possibly at the scale of a few nanometers or less (*i.e.* MRO scale).

From the MRO maps obtained from 4D-STEM (*e.g.* Fig 1b and 1c), we determined the average MRO domain size as a function of the scattering vector, $k$, for $Zr_{55}Co_{25}Al_{20}$ ($x = 1$), $Zr_{50}Cu_{25}Co_{12.5}Al_{12.5}$ ($x = 0.5$), and $Zr_{45}Cu_{50}Al_5$ ($x = 0$) (Fig. 3a). We also calculated the angular correlation power spectrum [28,34] from individual nanodiffraction patterns that are averaged over the entire sample (Fig. 3b). The *y*-axis of the power spectrum is the frequency ($n$) of the Fourier components of the angular correlation calculated from individual nanodiffraction patterns, and it represents the *n*-fold rotational symmetry present in the pattern. However, it is important to note that the higher order terms ($n = 4, 6, ..$) can always be created by the Fourier series (unless the angular correlation is perfectly sinusoidal, which is unlikely), so a particular $n$ value may not necessarily represent *n*-fold symmetry [14]. Therefore, to be safe, we do not differentiate between the even number *n*'s, but use them all together to indicate how strict the ordering is within the MRO domains depending on their power spectrum intensity. Odd number *n*'s in the spectrum may be artifacts due to the plural scattering when the sample is too thick [13,28]. While we use thin TEM samples with electron transmittance [35] of at least 60% to prevent such artifacts, there may still be some contribution from plural scattering, and therefore we excluded all odd *n*'s from our analysis. The trends both in the MRO size graph and power spectrum show a clear correlation to glass forming ability and ductility of the MGs as explained below.

First, $Zr_{55}Co_{25}Al_{20}$ ($x = 1$) shows relatively larger MRO size, about 1.2 nm at its peak, within the $k \sim 3.7$ to 4.1 nm$^{-1}$ (the blue curve in Fig. 3a). The corresponding power spectrum (top, Fig. 3b) shows high intensity of even-numbered $n$ values within that $k$ range, indicating that the MRO has high degree of atomic ordering that strongly diffracts the electrons. This suggests that the MRO in that $k$ range may be close to the nuclei of some crystalline phases (*e.g.* $\alpha$-Zr), which is consistent with the fact that this alloy, $Zr_{55}Co_{25}Al_{20}$, has low glass forming ability. There is another small peak at $\sim 5$ nm$^{-1}$, which indicates smaller interatomic spacing within the MRO. This peak likely corresponds to an MRO type that consists of mostly smaller atoms (*i.e.* Co), and the size of the MRO is $\sim 1$ nm.

Second, $Zr_{45}Cu_{50}Al_5$ ($x = 0$, red curve in Fig. 3a) shows a larger MRO size ($\sim 1.3$ nm) within the $k \sim 4.1$ to 4.6 nm$^{-1}$. However, the $n$'s in the corresponding power spectrum (bottom, Fig. 3b) show much lower intensity as compared to those in $Zr_{55}Co_{25}Al_{20}$, indicating that the dominant MRO in $Zr_{45}Cu_{50}Al_5$ and $Zr_{55}Co_{25}Al_{20}$ are different in terms of both their type and degree of order. The low power spectrum amplitude suggests that the dominant MRO in $Zr_{45}Cu_{50}Al_5$ should be more structurally frustrated (*i.e.* less ordered), which connects well with the high glass forming ability of that MG. Since the MRO is more structurally frustrated, it may contain more icosahedral SRO clusters widely observed in MD simulations (*e.g.* [5]). However, more details, such as, *e.g.*, how the icosahedral clusters are populated or aligned within the MRO, are not clear at this point. $Zr_{45}Cu_{50}Al_5$ also has a peak at $\sim 5$ nm$^{-1}$ that should correspond to the MRO mainly consist of Cu atoms, and the size of that MRO is about $\sim 1$ nm, almost the same as that of the Co-rich MRO in $Zr_{55}Co_{25}Al_{20}$.

Lastly, the "mixed" composition at $x = 0.5$, $Zr_{50}Cu_{25}Co_{12.5}Al_{12.5}$ (green curve in Fig. 3a), shows consistently smaller MRO size throughout the wider range of $k \sim 3.7$ to 4.6 nm$^{-1}$, which indicates more diverse distribution of MRO types, while showing the same MRO size ($\sim 1$ nm) for the one at $k \sim 5$ nm$^{-1}$. The MRO in $Zr_{50}Cu_{25}Co_{12.5}Al_{12.5}$ MG also appears to be less ordered than that of $Zr_{55}Co_{25}Al_{20}$ (middle, Fig.

3b). Importantly, the smaller and more diverse MRO in this $Zr_{50}Cu_{25}Co_{12.5}Al_{12.5}$ MG may correlate to its significantly higher ductility shown in Fig. 2b.

To summarize, the following conclusions can be drawn from the MRO size and angular correlation results above : (*i*) the glass forming ability of an MG may be directly connected to the size and distribution of certain types of MRO that are more structurally frustrated, (*ii*) the structurally frustrated MRO may contain more icosahedral SRO clusters, (*iii*) smaller size and more diverse distribution of MRO domains can substantially enhance ductility, and (*iv*) the MRO predominantly made with larger atoms (*e.g.* Zr) influences ductility more than the MRO made with mostly smaller atoms (*e.g.* Cu) does.

In addition, Fig. 3c shows the average area fraction of MRO appeared within the MRO maps. $Zr_{45}Cu_{50}Al_5$ (red curve) shows a high peak at $k \sim 4.3$ nm$^{-1}$, consistent with the peak in the size graph (red curve, Fig. 3a), indicating that a larger size of MRO domains directly leads to a larger area fraction in this MG. However, the $Zr_{55}Co_{25}Al_{20}$ (blue curve) displays a different trend, showing a lower fraction of MRO for $k \sim 4$ nm$^{-1}$ at which the MRO size is larger. The results again suggest that the dominant MROs in these two MGs have fundamentally different characteristics. The MRO in $Zr_{45}Cu_{50}Al_5$, which is more structurally frustrated, may be more related to the glass transition itself, likely inheriting the original composition (and perhaps some structure too) from the liquid state. Meanwhile the MRO in $Zr_{55}Co_{25}Al_{20}$ may be nucleated from the matrix perhaps shortly following (or during) the glass transition, given that their number per volume is limited possibly by the size of the diffusion field surrounding the MRO. This again matches well with our hypothesis that the dominant MRO in $Zr_{55}Co_{25}Al_{20}$ may be close to crystalline nuclei. Regardless of the type of MRO, the fact that the decrease in size of MRO (Fig. 3a) increases ductility (Fig. 2b) suggests that controlling the size of MRO may be the key to achieve tunable ductility in MGs.

We then used mesoscale deformation simulation to understand the detailed mechanism of the MRO-ductility correlation that we exprimentally observed above. Our mesoscale simulation maps MRO directly into different types of STZs that have different numbers of shearing mode, activation energy barrier, and softening behavior from the glassy matrix. The simulation can flexibly assume and integrate the important characteristics of the MRO to study its impact on the overall deformation beyond the spatial and temporal limits of MD simulations, which enables the consideration of real structural heterogeneities (*i.e.* the different MRO domains revealed by 4D-STEM), and simulations of their impact on the initiation and full development of multiple shear bands within one model [30–32]. This has allowed us to correlate the characterized structural heterogeneities and simulated properties directly to the experimentally measured properties of real MGs with comparable spatial and time scales to establish the direct microstructure-property relationships. Recently, we have used the simulation to capture several key MRO features revealed by fluctuation microscopy [31]. The study revealed that changing the volume fraction and type of MRO (*e.g.* 2-fold MRO vs. 6-fold MRO) domains has a significant effect on shear banding as well as on the stress-strain curves in tensile test simulations. Based on the result, we proposed the concept of "strain frustration" which is essentially related to the geometric incompatibility caused by dissimilar plastic carriers (*e.g.* different MROs) that exhibit strong bias in favor of certain local slip modes different from those of the glassy matrix. Since the local slip modes are greatly influenced by atomic packing, a correlation between MRO and the shear catalog of STZ should be expected naturally.

Based on the more detailed MRO parameters in MGs with different compositions that we experimentally determined using 4D-STEM, we aimed at establishing a constitutive description of the MRO-STZ relationship and offering a mechanistic understanding of the observed structure-property correlation in Figs. 2 and 3. The following heuristic rules about MRO-STZ relationship have been proposed: (*i*) the number of STZ shear modes is inversely proportional to the degree of ordering in the corresponding MRO domains, and (*ii*) the degree of softening introduced by STZ activation is inversely proportional to the degree of ordering in the corresponding MRO domains. The first rule is based on the argument that more ordering leads to more significant bias in favor of certain slip systems; the second rule reflects the fact that a more ordered atomic structure tends to preserve the original lattice sites (with crystals being the extreme case where lattice is completely preserved after the passage of a full dislocation) and hence less softening (*e.g.* free-volume or extended-defect based damage theory [30,36]). These theoretical rules allow the prescription of STZ properties for the glass systems that we experimentally examined above, $(Zr_{45}Cu_{50}Al_5)_{1-x}(Zr_{55}Co_{25}Al_{20})_x$ ($x$ = 0, 0.5, and 1). In particular, based on the measured degree of ordering (Fig. 3b), we consider the dissimilar STZs (derived from MROs) to have 12 shear modes in $Zr_{45}Cu_{50}Al_5$ ($x$ = 0) but 2 in $Zr_{55}Co_{25}Al_{20}$ ($x$ = 1), because the MRO in $Zr_{45}Cu_{50}Al_5$ is more disordered than that in $Zr_{55}Co_{25}Al_{20}$ as suggested by our experiments. Both cases have less shear modes as compared to the 20 shear modes of STZs derived from the glassy matrix, which is the most disordered region. Meanwhile, the glass of $x$ = 0.5 is considered to contain both types of the above dissimilar STZs with an equal population.

We then carry out tensile test simulations using these three MGs with the volume fraction of the dissimilar STZs all being 20%. While the peak stress as well as the stress-strain curves appear the same for all three glasses as shown in Fig. 4a (note that no damage model is included yet in the simulations), analysis on the largest connected-free-volume (CFV) [30] clearly shows that in the cases of $x$ = 0 and 1, an explosive growth of CFV up to ~100 nm³ occurs at a much earlier stage than that in the case of $x$ = 0.5 (Fig. 4b). The insets of Fig. 4b show the deformation microstructures (*i.e.*, von Mises strain maps) at the dashed line, with the largest CFV superimposed on top of them to better visualize the differences. It shows that a "run-away" shear band going through the entire sample is formed in both $x$ = 0 and 1 whereas the largest CFV in $x$ = 0.5 is still localized at the same level of macroscopic strain. Since the largest CFV is the most probable location for crack initiation, our results may imply that the mixture of glasses with different types of STZs (derived from different MROs) may actually provide improved ductility, consistent with the experimental result in Fig. 2b. Since the random number generator is used in the model (to account for the inherient randomness of the glass structure [32]), we have repeated the simulations for another six times with all different random number seeds. Figure 4c shows the average value of the threshold strain for percolation (defined as the strain with the largest slope the CFV vs. true strain plot as in Fig. 4b) for each composition, together with the error representing the standard deviation of each data set. It indeed confirms the statistical significance of the ductility trend observed in Fig. 4b. The "unconventional rule of mixture" above is likely attributed to the "strain frustration" [31] due to mixing different types of dissimilar STZs, since the 2-fold and 12-fold STZs are statistically incompatible with each other due to the mismatch of shear catalogs for accommodating local plastic shear. This incompatibility is likely to delay the percolation (growth) of connected free volume, since the local stress field is no longer concordant with the favorable shear catalog. This degree of strain frustration is obviously missing when the MG contains only one type of dissimilar STZs.

To consider the effect of MRO size, we first create additional two MRO maps, with one map contains much smaller MRO domains than the experimental value and the other one much larger MRO domains. A Gaussian distribution of the MRO domain size is also included in the synthesized MRO maps in order to be consistent with the experimental observation. Since the focus here is to identify unambiguously the effect of MRO domain size on shear banding, we have considered a significantly coarsened MRO structure, which may bear certain similarity as the glass containing nanocrystals. In all cases, the overall MRO volume fraction is kept the same to eliminate the possible influence of volume fraction. The result of this parametric study is shown in Fig. 4d, which clearly shows that increasing the MRO size can significantly decrease the ductility. The inset of Fig. 4d shows the deformation microstructures at 3% overall elongation for all three curves. It again shows that as MRO size increases, the shear band becomes much sharper and "hotter" (higher locally accumulated plastic strain), suggesting a more brittle glass. This simulation result is understandable as increasing the MRO size while keeping the volume fraction unchanged results in larger open regions in the absence of MRO, where shear bands can then run through easily. This argument is likely to apply to the current experiment as well, in that the volume fraction of the three samples are very close to each other (in particular, for those of $x = 0$ and $x = 0.5$) according to Fig. 3c. Interestingly, this rationale is similar to the loss of hardening effect as precipitates are sufficiently coarsened in crystalline materials, and we have in fact discussed this MRO-induced "precipitation-hardening" in our previous work [31] as well.

In summary, the quantitative analysis of MG structure using 4D-STEM clearly demonstrates the correlation between the detailed MRO parameters and important properties, including their ductility and glass forming ability. Mesoscale simulation based on the experimentally determined MRO information confirms that the diverse types and sizes of MRO domains can significantly influence the MGs' mechanical behavior. The new information we found here is critical as it provides important quantitative details of the structural heterogeneity in MGs and how it connects to their properties, which has been missing in the field. We believe the findings in this study may serve as an important foundation for establishing new paradigm in designing new amorphous materials with desired structural properties, for example, high strength combined with high ductility, by precise control of their nanoscale structures.

Acknowledgement

S.I., P.Z, Y. W and J. H. acknowledge support by the NSF under DMR-1709290. This work was performed in part at the Cornell PARADIM Electron Microcopy Facility, as part of the Materials for Innovation Platform Program, which is supported by the NSF under DMR-1539918 with additional infrastructure support from DMR-1719875 and DMR-1429155. D. M. and Z. C. are also supported by PARADIM.

[1]     E. Takeshi and S. J. L. Billinge, in *Underneath the Bragg Peaks*, edited by T. Egami and S. J. L. B. T.-P. M. S. Billinge (Pergamon, 2012), pp. 55–111.

[2]     C. E. Pueblo, M. Sun, and K. F. Kelton, Nat. Mater. **16**, 792 (2017).

[3]     D. Ma, A. D. Stoica, and X.-L. Wang, Nat. Mater. **8**, 30 (2009).

[4]     H. W. Sheng, W. K. Luo, F. M. Alamgir, J. M. Bai, and E. Ma, Nature **439**, 419 (2006).

[5]     Y. Q. Cheng, E. Ma, and H. W. Sheng, Phys. Rev. Lett. **102**, 245501 (2009).

[6]     D. B. Miracle, Nat. Mater. **3**, 697 (2004).

[7]     J. M. Rodenburg, Ultramicroscopy **25**, 329 (1988).

[8]     M. M. J. Treacy, J. M. Gibson, L. Fan, D. J. Paterson, and I. McNulty, Reports Prog. Phys. **68**, 2899 (2005).

[9]     J. Hwang, Z. H. Melgarejo, Y. E. Kalay, I. Kalay, M. J. Kramer, D. S. Stone, and P. M. Voyles, Phys. Rev. Lett. **108**, 195505 (2012).

[10]    P. M. Voyles and D. A. Muller, Ultramicroscopy **93**, 147 (2002).

[11]    J. J. Maldonis, J. Hwang, and P. M. Voyles, Comput. Phys. Commun. **213**, 217 (2017).

[12]    J. Hwang, A. M. Clausen, H. Cao, and P. M. Voyles, J. Mater. Res. **24**, 3121 (2009).

[13]    A. C. Y. Liu, M. J. Neish, G. Stokol, G. A. Buckley, L. A. Smillie, M. D. de Jonge, R. T. Ott, M. J. Kramer, and L. Bourgeois, Phys. Rev. Lett. **110**, 205505 (2013).

[14]    T. Sun, M. M. J. Treacy, T. Li, N. J. Zaluzec, and J. Murray Gibson, Microsc. Microanal. **20**, 627 (2014).

[15]    Y. E. Kalay, I. Kalay, J. Hwang, P. M. Voyles, and M. J. Kramer, Acta Mater. **60**, (2012).

[16]    Y. H. Liu, D. Wang, K. Nakajima, W. Zhang, A. Hirata, T. Nishi, A. Inoue, and M. W. Chen, Phys. Rev. Lett. **106**, 125504 (2011).

[17]    Y. H. Liu, G. Wang, R. J. Wang, D. Q. Zhao, M. X. Pan, and W. H. Wang, Science

(80-. ). **315**, 1385 LP (2007).

[18] B. Zhang, D. Q. Zhao, M. X. Pan, W. H. Wang, and A. L. Greer, Phys. Rev. Lett. **94**, 205502 (2005).

[19] J. M. Park, J. H. Han, N. Mattern, D. H. Kim, and J. Eckert, Metall. Mater. Trans. A **43**, 2598 (2012).

[20] A. S. Argon, Acta Metall. **27**, 47 (1979).

[21] A. Widmer-Cooper and P. Harrowell, J. Non. Cryst. Solids **352**, 5098 (2006).

[22] M. L. Manning and A. J. Liu, Phys. Rev. Lett. **107**, 108302 (2011).

[23] J. Ding, S. Patinet, M. L. Falk, Y. Cheng, and E. Ma, Proc. Natl. Acad. Sci. **111**, 14052 (2014).

[24] J. Ding, Y.-Q. Cheng, H. Sheng, M. Asta, R. O. Ritchie, and E. Ma, Nat. Commun. **7**, 13733 (2016).

[25] F. Spaepen, Acta Metall. **25**, 407 (1977).

[26] M. L. Falk, J. S. Langer, and L. Pechenik, Phys. Rev. E **70**, 11507 (2004).

[27] Y. Fan, T. Iwashita, and T. Egami, Nat. Commun. **8**, 15417 (2017).

[28] S. Im, Z. Chen, J. M. Johnson, P. Zhao, G. H. Yoo, E. S. Park, Y. Wang, D. A. Muller, and J. Hwang, Ultramicroscopy **198**, 189 (2018).

[29] M. W. Tate, P. Purohit, D. Chamberlain, K. X. Nguyen, R. Hovden, C. S. Chang, P. Deb, E. Turgut, J. T. Heron, D. G. Schlom, D. C. Ralph, G. D. Fuchs, K. S. Shanks, H. T. Philipp, D. A. Muller, and S. M. Gruner, Microsc. Microanal. **22**, 237 (2016).

[30] P. Zhao, J. Li, and Y. Wang, Acta Mater. **73**, 149 (2014).

[31] P. Zhao, J. Li, J. Hwang, and Y. Wang, Acta Mater. **134**, 104 (2017).

[32] P. Zhao, J. Li, and Y. Wang, Int. J. Plast. **40**, 1 (2013).

[33] J. Hwang and P. M. Voyles, Microsc. Microanal. **17**, 67 (2011).

[34] P. Wochner, C. Gutt, T. Autenrieth, T. Demmer, V. Bugaev, A. D. Ortiz, A. Duri, F. Zontone, G. Grübel, and H. Dosch, Proc. Natl. Acad. Sci. **106**, 11511 (2009).

[35] D. T. Schweiss, J. Hwang, and P. M. Voyles, Ultramicroscopy **124**, (2013).

[36] L. Li, E. R. Homer, and C. A. Schuh, Acta Mater. **61**, 3347 (2013).

Figure Captions

Figure 1. (a) Schematic of 4D-STEM. (a) Nanodiffraction patterns are acquired using electron probe (diameter = 1 nm) from oversampled probe positions ($p_1$, $p_2$, ..) on the sample. The intensities ($i_1$, $i_2$ ,..) in the acquired stack of patterns can then be reconstructed in the real space by selecting any ($k_x$, $k_y$) pixel within the pattern. (b and c) The reconstructed "dark-field" using the "b" and "c" pixels in (a), respectively.

Figure 2. Experimental (a) X-ray diffraction, (b) compression, (c) DSC, and (d) atom probe tomography data from $(Zr_{45}Cu_{50}Al_5)_{1-x}(Zr_{55}Co_{25}Al_{20})_x$ ($x$ = 0, 0.5, and 1).

Figure 3. (a) Average MRO size (see Supplemental Materials for the same data with error bars) (b) average power spectrum calculated from each nanodiffraction pattern, and (c) area fraction of MRO as a function of $k$ determined from $Zr_{55}Co_{25}Al_{20}$ ($x$ = 1), $Zr_{50}Cu_{25}Co_{12.5}Al_{12.5}$ ($x$ = 0.5), and $Zr_{45}Cu_{50}Al_5$ ($x$ = 0) using 4D-STEM.

Figure 4. (a) Simulated stress-strain curves for $(Zr_{45}Cu_{50}Al_5)_{1-x}(Zr_{55}Co_{25}Al_{20})_x$ ($x$ = 0, 0.5, and 1). (b) Evolution of the largest connective-free-volume (CFV) during deformation. Inset shows the Von Mises strain maps (red dots) at the dashed line in (b) and superposition of the largest CFV (cyan dots) on top of them. (c) Average threshold strain for percolation for each composition, together with the error representing the standard deviation. (d) Simulated stress-strain curves varying MRO size. The inset shows the deformation microstructures at 3% overall elongation for all three curves.

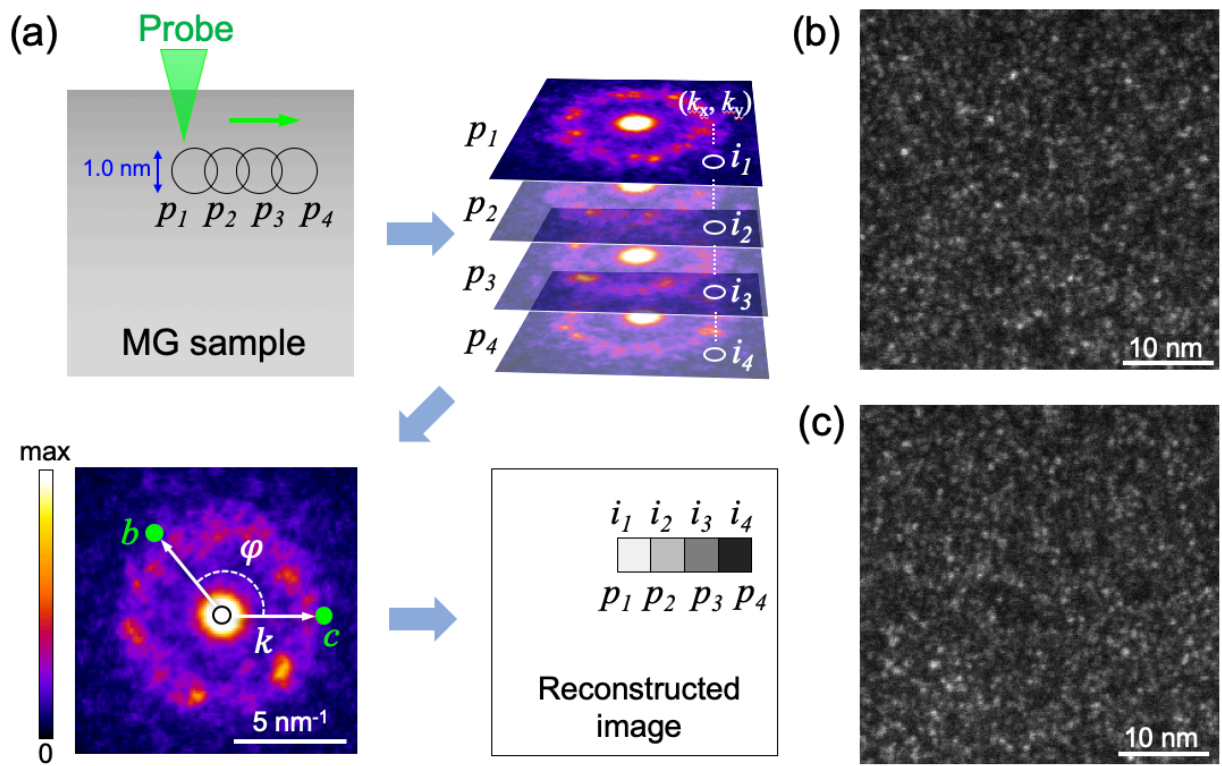

Figure 1. (a) Schematic of 4D-STEM. (a) Nanodiffraction patterns are acquired using electron probe (diameter = 1 nm) from oversampled probe positions ($p_1$, $p_2$, ..) on the sample. The intensities ($i_1$, $i_2$ ,..) in the acquired stack of patterns can then be reconstructed in the real space by selecting any ($k_x$, $k_y$) pixel within the pattern. (b and c) The reconstructed "dark-field" using the "b" and "c" pixels in (a), respectively.

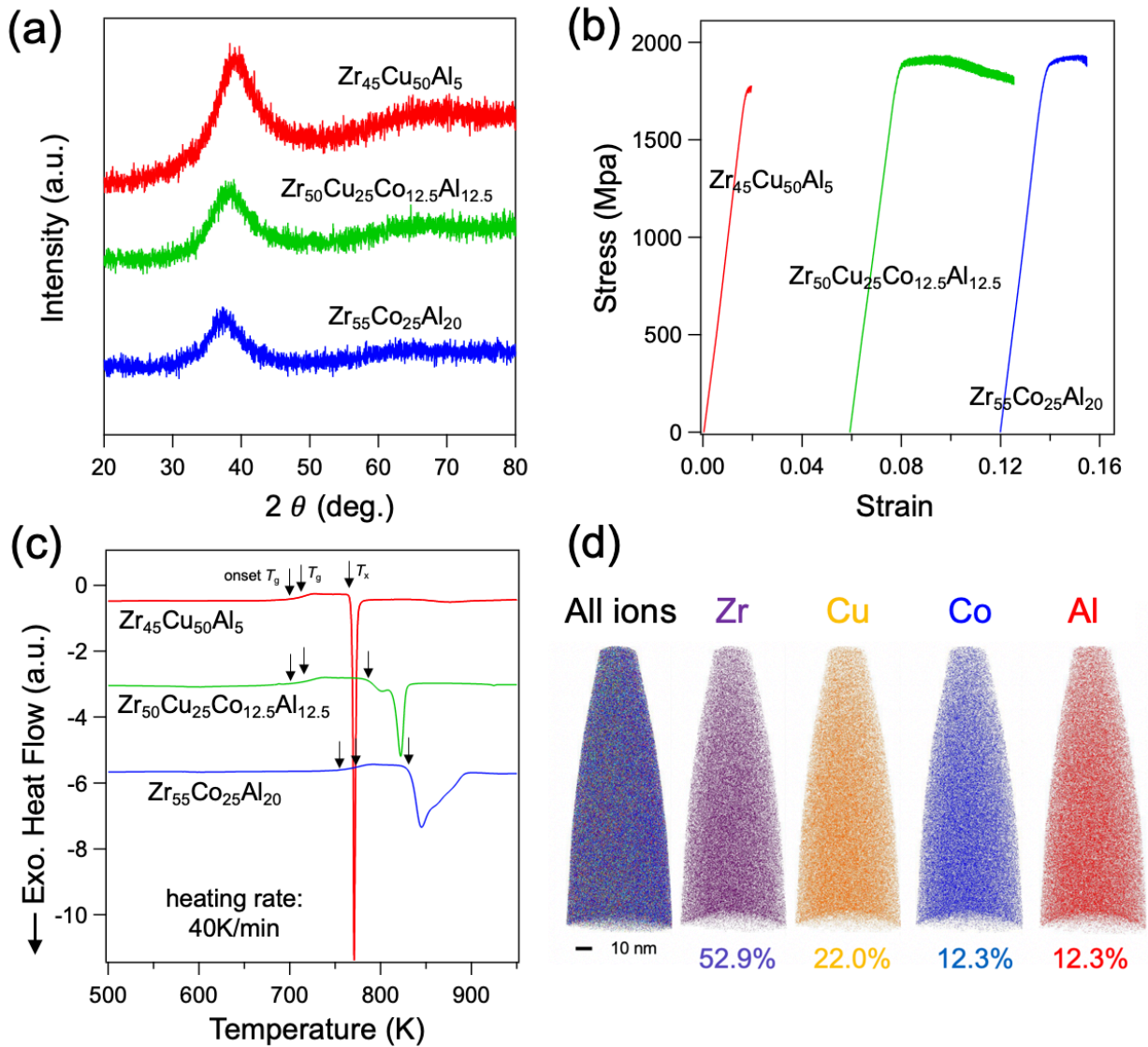

Figure 2. Experimental (a) X-ray diffraction, (b) compression, (c) DSC, and (d) atom probe tomography data from $(Zr_{45}Cu_{50}Al_5)_{1-x}(Zr_{55}Co_{25}Al_{20})_x$ ($x$ = 0, 0.5, and 1).

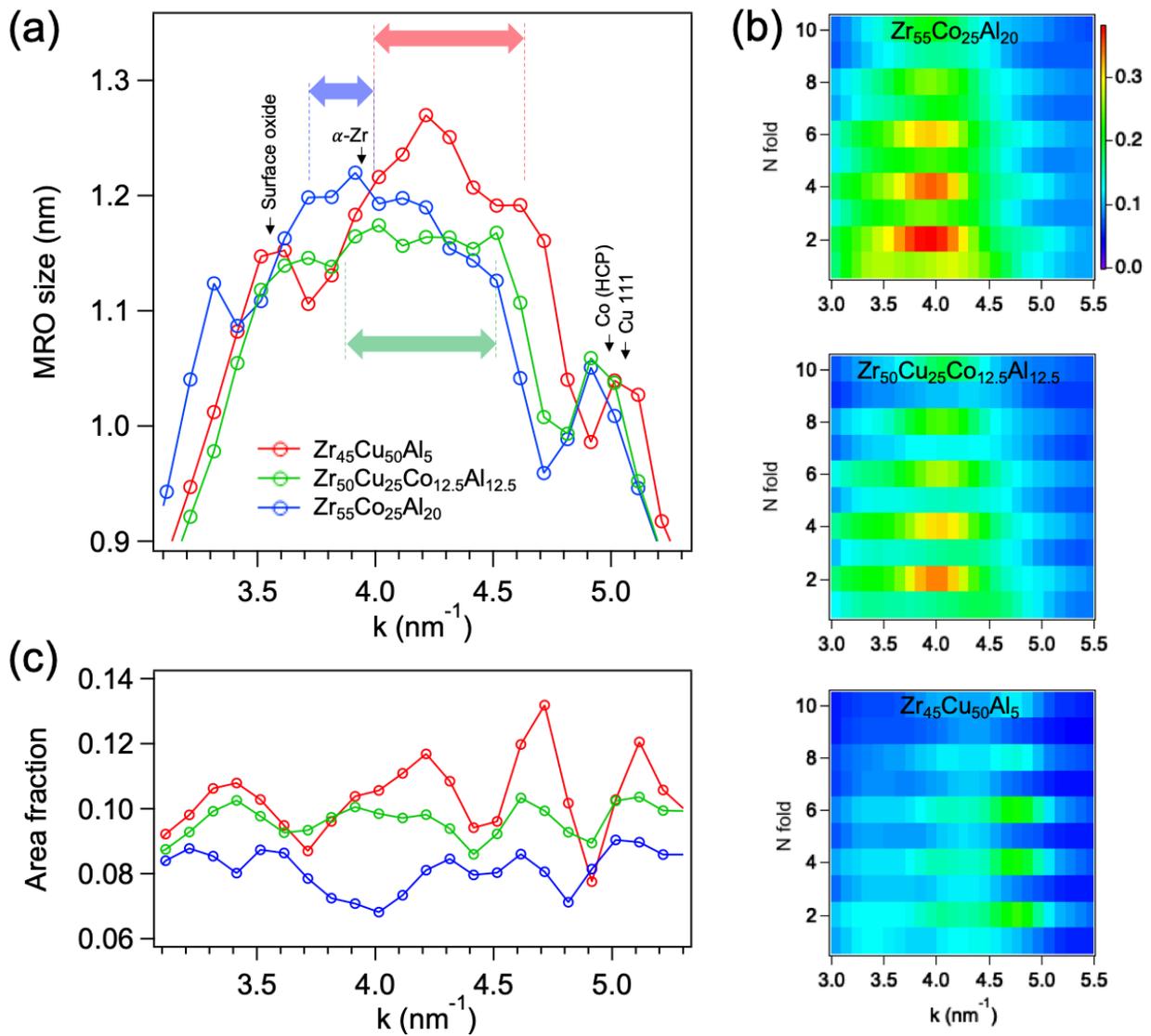

Figure 3. (a) Average MRO size (see Supplemental Materials for the same data with error bars) (b) average power spectrum calculated from each nanodiffraction pattern, and (c) area fraction of MRO as a function of $k$ determined from $Zr_{55}Co_{25}Al_{20}$ ($x = 1$), $Zr_{50}Cu_{25}Co_{12.5}Al_{12.5}$ ($x = 0.5$), and $Zr_{45}Cu_{50}Al_5$ ($x = 0$) using 4D-STEM.

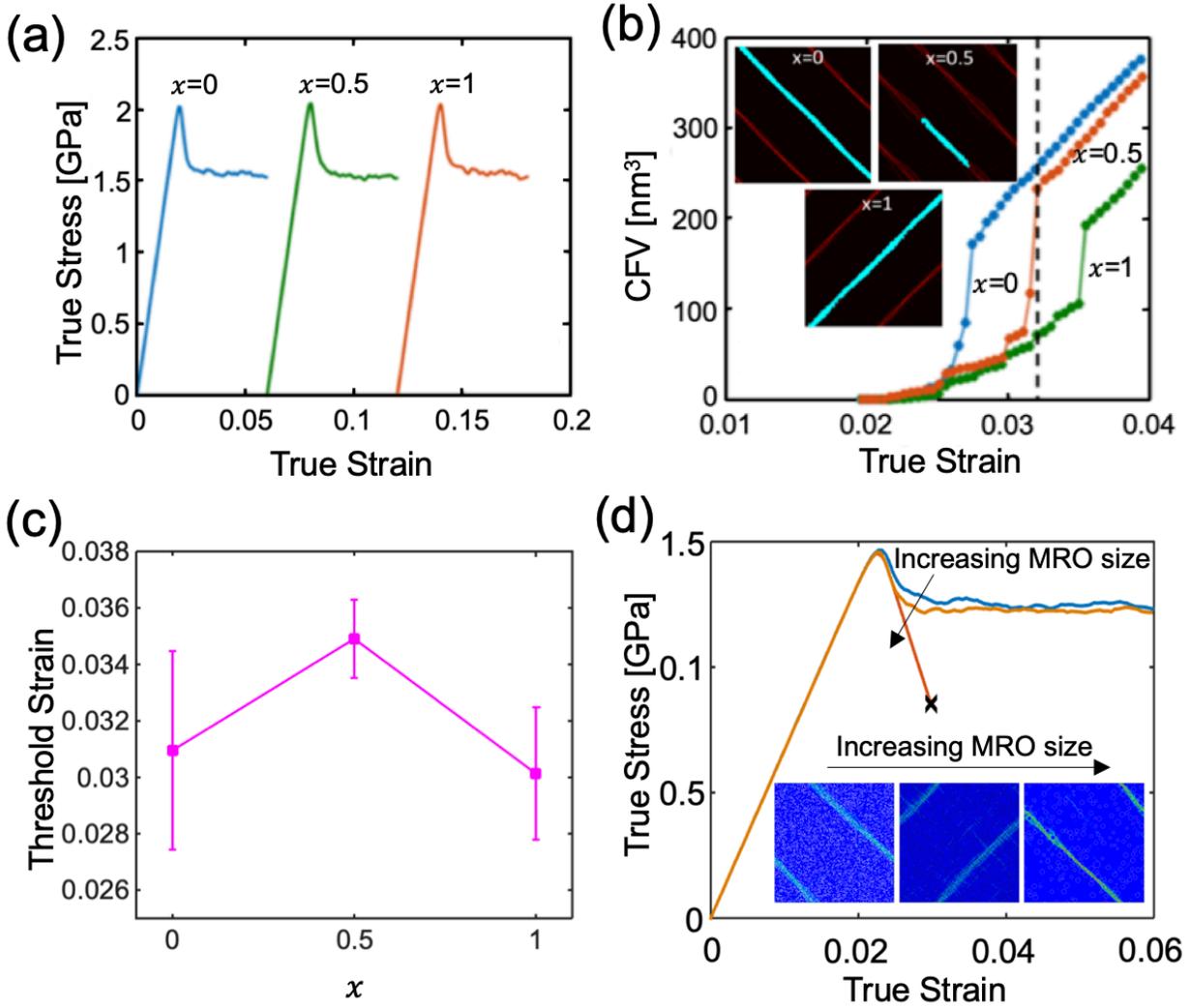

Figure 4. (a) Simulated stress-strain curves for $(Zr_{45}Cu_{50}Al_5)_{1-x}(Zr_{55}Co_{25}Al_{20})_x$ ($x$ = 0, 0.5, and 1). (b) Evolution of the largest connective-free-volume (CFV) during deformation. Inset shows the Von Mises strain maps (red dots) at the dashed line in (b) and superposition of the largest CFV (cyan dots) on top of them. (c) Average threshold strain for percolation for each composition, together with the error representing the standard deviation. (d) Simulated stress-strain curves varying MRO size. The inset shows the deformation microstructures at 3% overall elongation for all three curves.